\documentclass{article}
\usepackage{ijcai24}

\usepackage{times}
\usepackage{soul}
\usepackage{url}
\usepackage[hidelinks]{hyperref}
\usepackage[utf8]{inputenc}
\usepackage[small]{caption}
\usepackage{graphicx}
\usepackage{amsmath}
\usepackage{amsthm}
\usepackage{booktabs}
\usepackage{algorithm}
\usepackage{algorithmic}
\usepackage{bbding}
\usepackage{utfsym}
\usepackage[switch]{lineno}

\usepackage{multirow}
\usepackage{amssymb}

\newcommand{\paratitle}[1]{\vspace{1.5ex}\noindent\textbf{#1}}
\newcommand{\ie}{\emph{i.e.,}}

\newcommand{\fig}{Figure}

\title{Side Information-Driven Session-based Recommendation: A Survey}

\author{
Xiaokun Zhang$^{1}$
\and
Bo Xu$^1$\and
Chenliang Li$^{2}$\and
Yao Zhou$^3$\and
Liangyue Li$^4$\and
Hongfei Lin$^1$
\affiliations
$^1$Dalian University of Technology,
$^2$Wuhan University,
$^3$Instacart Inc,
$^4$Alibaba Group.\\
\emails
dawnkun1993@gmail.com,
xubo@dlut.edu.cn,
cllee@whu.edu.cn,\\
collwe01@gmail.com,
liliangyue.lly@alibaba-inc.com,
hflin@dlut.edu.cn
}

\begin{document}

\maketitle

\begin{abstract}
The session-based recommendation (SBR) garners increasing attention due to its ability to predict anonymous user intents within limited interactions. Emerging efforts incorporate various kinds of side information into their methods for enhancing task performance. In this survey, we thoroughly review the side information-driven session-based recommendation from a data-centric perspective. Our survey commences with an illustration of the motivation and necessity behind this research topic. This is followed by a detailed exploration of various benchmarks rich in side information, pivotal for advancing research in this field. Moreover, we delve into how these diverse types of side information enhance SBR, underscoring their characteristics and utility. A systematic review of research progress is then presented, offering an analysis of the most recent and representative developments within this topic. Finally, we present the future prospects of this vibrant topic.

\end{abstract}

\section{Introduction}


Recommender systems, as an indispensable tool to combat information overload, play a vital role in the current information era. Conventional recommender systems mainly rely on user profiles and long-term behaviors to model their preferences and predict future purchases or clicks. However, due to the stringent privacy policies, it is hard to access such sensitive identity information to conduct recommendations. 
To address this prevalent scenario, session-based recommendation (SBR) is proposed to predict the next items of interest for \textbf{anonymous} users within \textbf{short} sessions~\cite{GRU4Rec,NARM}. Owing to its significant application value, SBR has garnered widespread attention since its inception~\cite{GRU4Rec,NARM,Wang@CS2022}.

\begin{figure}[t]
  \centering
  \includegraphics[width=0.90\linewidth]{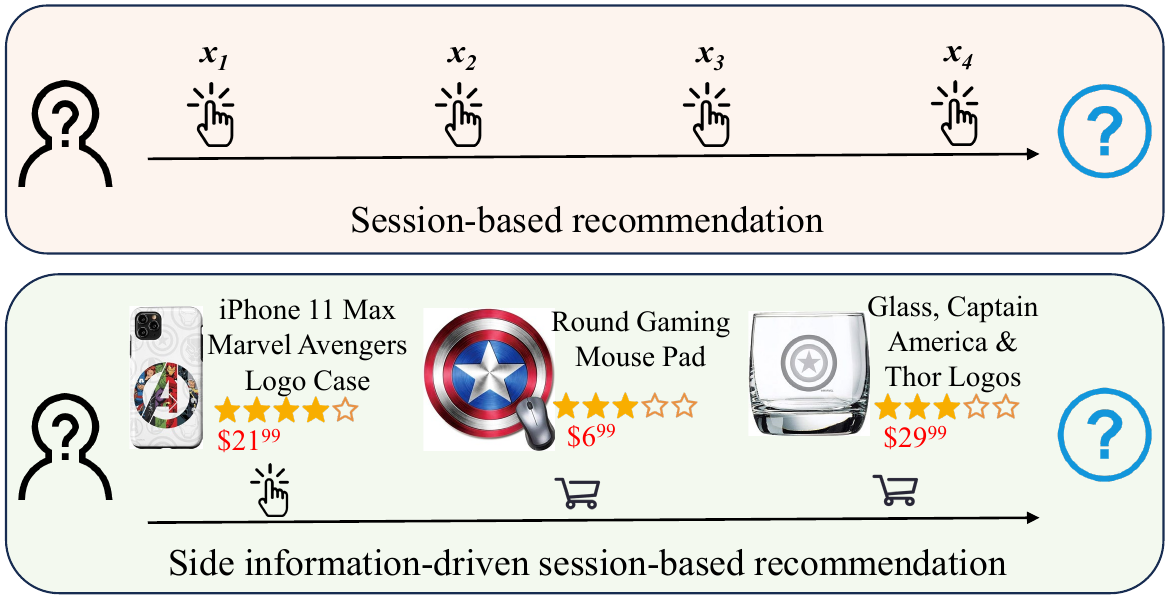}
  \caption{Conventional SBR v.s. side information-driven SBR.}\label{intro}
\end{figure}

However, SBR inevitably suffers from serious data sparsity issues~\cite{DHCN,Yang@SIGIR2023}. As illustrated in the first row of Figure~\ref{intro}, SBR typically relies on a limited number of user interactions, like a few clicks, to infer a user's intent, substantially increasing the task's difficulty.
To this end, as shown in the second row of Figure~\ref{intro}, there is a growing trend to incorporate various types of side information, such as item images, title text, price, and behavior types, to facilitate user intent understanding within the constraints of SBR. This survey aims to provide a comprehensive review of the benchmarks, data characteristics and utility, progress, and prospects in the vibrant field of side information-driven session-based recommendation.

\paratitle{Motivation: \textit{Why side information-driven SBR?}}

\paragraph{(1) Enriching Session Data.}
Side information enriches session data beyond mere item IDs, offering important clues into user behaviors. For instance, the categories of items a user interacts with can effectively narrow down the potential scope for their next actions, thereby enhancing prediction accuracy~\cite{xie@SIGIR2022}. In the current landscape, where neural models play a dominant role in this field, the adage ``more data leads to better performance'' typically holds true. The inclusion of side information presents a valuable opportunity to maximize the capabilities of neural networks, unlocking their vast potential in making more accurate recommendations.

\paragraph{(2) Depicting Item Detailed Features.}
Item ID, as a main indicator in current methods to handle SBR, merely serves as a symbolic identifier, simply suggesting item co-occurrence patterns in a statistical perspective~\cite{MMSBR}. In contrast, side information provides a vivid portrayal of various item features. For example, images can convey a clothing's style, such as whether it is tight or loose, while the price reflects its value. Side information, therefore, offers a comprehensive understanding of an item, encompassing factors that are crucial in determining users' choices.

\paragraph{(3) Revealing User Fine-grained Preferences.} 
Rich in semantic content, side information holds the ability to reveal user fine-grained preferences~\cite{MMMLP,Hu@CIKM2023}. As shown in the second row of Figure~\ref{intro}, we could infer that a user is a Marvel enthusiast based on the images associated with the items he interacted with. This insight opens up the possibility of making tailored suggestions, such as recommending a T-shirt featuring an Iron-Man logo to this Marvel fan. In essence, detailed insights obtained from side information empower models to provide highly personalized services, catering to the specific interests of individuals.


\paratitle{Paradigm: \textit{How to handle side information-driven SBR?}}

\noindent The emergence of side information-driven SBR has coincided with the rise of deep learning, leading to the dominance of neural models in this field. Generally, neural models focus on obtaining item and session embeddings, basing recommendations on their similarity. Let $\mathcal{I}$ ($|\mathcal{I}|=n$) denote unique item set. An item $x_i \in \mathcal{I}$ may contain various kinds of side information, including time, categories, price, title text, images, and so on. Formally, given a session ${s}$ = [$x_1, x_2, ..., x_m$] ($x_j \in \mathcal{I}$), which contains an anonymous user's actions within a short period, the score of a candidate item $x_i \in \mathcal{I}$ being the next interaction is predicted as,
\begin{align}
    P(s, x_i) &= {\rm sim}<f(s), g(x_i)>,
\end{align}
where $f(\cdot)$ is a neural network designed to represent sessions such as Recurrent Neural Network(RNN)~\cite{GRU4Rec}, attention mechanism~\cite{STAMP} or Graph Neural Network(GNN)~\cite{SR-GNN}, $g(\cdot)$ is used to obtain item embeddings, and ${\rm sim}<\cdot>$ evaluates their similarity like cosine function. Finally, the items with top-$k$ scores are formulated as the recommendation list.

\paratitle{Novelty: \textit{What are the differences between this survey and previous ones?}}

\noindent This survey distinguishes itself from existing works in several key areas, including reviews on sequential recommendation(SR)~\cite{Wang@IJCAI2019}, multi-modal recommendation(MMRec)~\cite{Zhou@arXiv2023} and session-based recommendation(SBR)~\cite{ludewig@UMUAI2020}. Firstly, we want to clarify the distinction between two commonly confused concepts, SR and SBR. For SBR, where the users are anonymous, it aggregates item embeddings within a session to obtain the session embeddings. Conversely, SR assumes the availability of user profiles, utilizing user-specific information (like user ID or demographics) to learn individualized user embeddings. This survey is specifically tailored to explore various approaches that are uniquely suited to the task of SBR. 
Secondly, requiring user profiles, MMRec's direct applicability to SBR is often limited. Moreover, different from current MMRec mainly focusing on text and images, this survey extends the scope to include extra information particularly relevant to SBR, e.g., time which can shed light on user dynamic intent within a session. Thirdly, existing reviews on SBR mainly focus on a single kind of information, i.e., item ID, and summarize related works in a technique view. In contrast, our survey embraces a wider range of data types and clarifies this topic from a data-centric perspective.

\paratitle{Contributions:} The main contributions of this survey are summarized as follows:
\begin{itemize}
    \item We comprehensively outline available benchmarks for side information-driven SBR, which is crucial for further research in this field.
    
    \item We systematically analyze the characteristics and utility of various kinds of side information, providing new perspectives to incorporate side information into SBR. 
    
    \item We summarize the current research progress by categorizing studies according to the types of information they utilize, aiding in understanding the existing landscape and offering convenience for future research in replicating, comparing, and enhancing these methods.

    \item We highlight and discuss several promising research directions for this topic, looking forward to inspiring the community.
    
\end{itemize}

\section{Benchmarks}
Datasets, as the collections of user-item interaction data, are the foundation for the research of recommender systems. They serve as the essential groundwork for designing, training, and evaluating recommendation algorithms. This is particularly true for side information-driven SBR, where the types of information available in a dataset dictate the range of research questions it supports. For instance, studying user price sensitivity is infeasible, with datasets lacking price information. Therefore, this section provides a detailed introduction to the available datasets.

We present a list of available datasets, along with their associated information types, in Table~\ref{datasets}. These datasets are selected from papers on SBR since its inception in 2016~\cite{GRU4Rec}. Besides, we have scrutinized works of conventional recommendation in the last five years, including the datasets from these works appropriate to our researched topic. 
Relying on different vehicles for conveying particular item features, various kinds of side information are categorized into two groups: \textbf{numerical information} of $\heartsuit$, and \textbf{descriptive information} of $\spadesuit$~\cite{MMSBR}. The numerical information delivers the abstract meaning of an item using numbers,  such as integer IDs for the category or real numbers for the item price. 
Descriptive information characterizes an item via text or images which could encompass rich item semantics such as color, style, and so on. The detailed introduction of these datasets is outlined as follows,

\begin{table*}[th]
\small
\tabcolsep 0.11in 
  \centering

\begin{tabular}{c|c|c|c|    c|c|c|c|    c|c|c}
\toprule
{Dataset}  & {Time$^*$} & {Category}    & {Brand}   & {Price}   & {Title}   & {Description} & {Image}       & {Rating$^*$}  & {Review$^*$}  & {Behavior$^*$}\\ 
\hline
{Amazon} & {$\heartsuit$} & {$\spadesuit$}  & {$\spadesuit$}  & {$\heartsuit$}  & {$\spadesuit$}  &{$\spadesuit$}   & {$\spadesuit$}    & {$\heartsuit$}  &{$\spadesuit$}   & {}  \\ 
 \hline
{Beer} & {$\heartsuit$}  & {$\spadesuit$}                     & {} & {} & {$\spadesuit$}                     & {} & {}    & {$\heartsuit$}  & {$\spadesuit$} & {} \\ 
\hline
{Cosmetics} &{$\heartsuit$} & {$\heartsuit$}  & {$\spadesuit$}  & {$\heartsuit$}  & {}    & {}    & {}       & {}    & {}    & {$\spadesuit$}  \\ 
\hline
{Diginetica} & {$\heartsuit$}   & {$\heartsuit$}  & {}    & {$\heartsuit$}  & {$\heartsuit$}  & {}    & {}      &{} & {}    & {}    \\
\hline
{Goodreads}  & {$\heartsuit$} &{} &{} &{} &{$\spadesuit$}   & {$\spadesuit$}  & {$\spadesuit$}    & {$\heartsuit$}  & {$\spadesuit$}  & {}    \\ 
\hline
{Instacart} & {$\heartsuit$}  &{$\spadesuit$} &{} &{} &{$\spadesuit$}   & {}  & {}    & {}  & {}  & {$\spadesuit$}    \\ 
\hline
{JingDong} & {$\heartsuit$} & {$\heartsuit$}                     & {} & {} & {} & {} & {}       & {} & {} & {$\spadesuit$}   \\ 
\hline
{LastFM}    & {$\heartsuit$}   & {$\spadesuit$}   & {} & {} & {$\spadesuit$}    & {}    & {}     & {}  & {}  & {}    \\ 
\hline
{Movielens}   & {$\heartsuit$}   & {$\spadesuit$}   & {} & {} & {$\spadesuit$}    & {}    & {}      & {$\heartsuit$}  & {$\spadesuit$}  & {}    \\ 
\hline
{RetailRocket} & {$\heartsuit$}  & {$\heartsuit$}   & {} & {} & {} & {} & {}       & {} & {} & {$\spadesuit$}  \\ 
\hline
{Steam}  & {$\heartsuit$}  & {$\spadesuit$}  & {}    & {$\heartsuit$}  & {$\spadesuit$}  & {} & {}    & {} & {$\spadesuit$} & {}    \\ 
\hline
{Taobao}     & {$\heartsuit$}    & {$\heartsuit$}                     & {} & {} & {} & {} & {}       & {} & {} & {$\spadesuit$}                 \\ 
\hline
{Tmall}       & {$\heartsuit$}     & {$\heartsuit$}                     & {$\heartsuit$}                     & {} & {} & {} & {}     & {} & {} & {$\spadesuit$}  \\ 
\hline
{Trivago}   & {$\heartsuit$}     & {} & {} & {$\heartsuit$}                     & {} & {} & {}       & {} & {} & {$\spadesuit$} \\ 
\hline
{Yelp}       & {$\heartsuit$}    & {$\spadesuit$}                     & {} & {} & {} & {} & {}       & {$\heartsuit$}                     & {$\spadesuit$}  & {}   \\ 
\hline
{Yoochoose}   & {$\heartsuit$}    & {$\heartsuit$}                     & {} & {$\heartsuit$}                     & {} & {} & {}      & {} & {} & {}  \\
\bottomrule
\end{tabular}
\caption{Available datasets for side information-driven SBR. The $^*$ denotes \textbf{user-item interaction} information, while other types of information without this symbol belong to items' \textbf{intrinsic properties}. The $\heartsuit$ indicates the information is presented by \textbf{numerical} format, and the $\spadesuit$ is for \textbf{descriptive} format, while the blank means the relevant information is absent. .}
\label{datasets}
\end{table*}

\begin{itemize}
    
    \item \textbf{Amazon}\footnote{\url{http://jmcauley.ucsd.edu/data/amazon/}} is a popular dataset scratched from the E-commerce website Amazon. It contains many sub-datasets covering various domains such as beauty, cell phones, and so on. Typically, each sub-dataset is treated as a distinct dataset for recommendation tasks.
    
    \item \textbf{Beer}\footnote{\url{https://cseweb.ucsd.edu/$\sim$jmcauley/datasets.html\#multi\_aspect}} contains user reviews about beer from Ratebeer and Beeradvocate. Besides the information listed in the Table, it records sensory aspects such as taste, look, feel, and smell.

    \item \textbf{Cosmetics}\footnote{\url{https://www.kaggle.com/mkechinov/ecommerce-events-history-in-cosmetics-shop}} records behavior data for 5 months in a medium cosmetics online store. Generally, each month's data can be used as a unique dataset.

    \item \textbf{Diginetica}\footnote{\url{https://competitions.codalab.org/competitions/11161}} is a competition dataset used in CIKM Cup 2016, recording user sessions extracted from an e-commerce search engine logs.

    \item \textbf{Goodreads}\footnote{\url{https://mengtingwan.github.io/data/goodreads}} contains users' reviews towards books from the Goodreads book review website.

    \item \textbf{Instacart}\footnote{\url{https://www.kaggle.com/c/instacart-market-basket-analysis}} is an anonymized dataset containing grocery orders from users on the online grocery marketplace Instacart. Owing to the vast volume of data within this dataset, it is a common practice to sample a subset of the data for research purposes.

    \item \textbf{JingDong}\footnote{\url{https://tinyurl.com/ybo8z4yz}} is extracted from JD.com, which is a famous Chinese e-commerce website. Similar to the Amazon dataset, this dataset also contains two product categories, ``Appliances'' and ``Computers'', commonly used as separate datasets.


    \item \textbf{LastFM}\footnote{\url{http://ocelma.net/MusicRecommendationDataset/index.html}} is a music-artist dataset specially designed for music interest recommendation. In this dataset, the tags of the artists given by users are usually viewed as the category information.

    \item \textbf{Movielens}\footnote{\url{https://grouplens.org/datasets/movielens/}} is a classical rating dataset about movies from the MovieLens website. Data with different volumes of Movielens often forms different datasets like Movielens-100k or Movielens-1m.

    \item \textbf{RetailRocket}\footnote{\url{https://www.kaggle.com/retailrocket/ecommerce-dataset}} is generated from an online shopping site-Retailrocket over a period of 4.5 months.

    \item \textbf{Steam}\footnote{\url{https://github.com/kang205/SASRec/tree/master/data}} is crawled from Steam, a large online video game distribution platform, containing users' reviews about games.

    \item \textbf{Taobao}\footnote{\url{https://tianchi.aliyun.com/dataset/dataDetail?dataId=649}} records users' online shopping behaviors from Taobao, a large e-commerce platform in China.

    \item \textbf{Tmall}\footnote{\url{https://tianchi.aliyun.com/dataset/dataDetail?dataId=42}} is user-purchase data obtained from Tmall (the largest B2C platform in China) and used in IJCAI 2015 competition.

    \item \textbf{Trivago}\footnote{\url{http://www.recsyschallenge.com/2019/}} is provided by Trivago, a global hotel search platform, recording user actions on the hotel.

    \item \textbf{Yelp}\footnote{\url{https://www.yelp.com/dataset}} captures users' reviews for restaurants from the online review platform Yelp.com. Due to its substantial size, existing works often utilize only a part of this dataset in their study.

    \item \textbf{Yoochoose}\footnote{\url{http://2015.recsyschallenge.com/challege.html}} contains click-streams from an E-commerce website within 6 months and is used as a challenge dataset for RecSys Challenge 2015.
    
\end{itemize}

\section{Side Information Characteristics and Utility}
Side information-driven SBR is centered around the strategic use of diverse data to enhance user intent capturing in the context of SBR. Each type of information possesses its unique characteristics and can shed light on different behavior patterns of users, thereby enhancing SBR from different perspectives. 
In the subsequent sub-sections, we will delve into the characteristics of the currently available information and highlight their specific utility in enhancing SBR.

\subsection{Time Information}
Time information records the precise timestamp of user-item interactions.  This information is usually presented in the form of UNIX time, which can be readily converted into the ``year-month-hour-minute-second'' format, providing a granular view of user activities. In the context of SBR, time information is crucial for identifying the sequential dependencies among items—a key aspect of modeling the dynamic nature of user behaviors. Varying granularity of time information allows for the detection of various sequential patterns in user behaviors, significantly enhancing recommendation performance. For example, by discerning time at the day-level granularity, it becomes logical to recommend fast food on weekdays when users might prefer quicker meal options, and suggest more substantial meals on weekends when they may have more time to enjoy leisurely dining. Similarly, time information can be leveraged to capture seasonal trends, enabling the nice suggestions of thick-down jackets during winter and lightweight T-shirts in summer. 

\subsection{Category Information}
{Category} information refers to the hierarchical classification or taxonomy used to organize and group items based on their types, characteristics, or other attributes. This category information suggests a user hierarchical decision-making process, such as navigating from ``Clothes'' to ``Men's Clothing'' and further to ``Pants''. These hierarchical patterns offer a strategic advantage by enabling methods to incorporate them to narrow down the range of candidate items effectively, thereby enhancing the effectiveness of recommendations. For instance, it will be wise to suggest cases in the sub-category ``phone-accessories'' to users who have recently purchased a cellphone in the category ``Smartphones''.

\subsection{Brand Information}
{Brand} information encapsulates specific details and attributes associated with the manufacturer or creator of an item.
The brand of an item is a critical factor in guiding customers' purchasing decisions, as it often conveys quality, reputation, and other intangible values associated with a particular brand. The ``brand effect'', as documented in economic research, underscores the impact of brand loyalty on consumer behavior. This phenomenon can be leveraged by recommender systems to offer more personalized and relevant suggestions. For instance, a user with a strong affinity for a specific brand tends to respond positively to recommendations featuring items from that brand. This alignment with user preferences can significantly enhance the effectiveness of recommendation strategies, making `Brand' information a valuable asset to meet individual tastes. 

\subsection{Price Information}
{Price} refers to the monetary cost assigned to an item, representing the amount a user is expected to pay for its acquisition.  It stands as a pivotal factor in the decision-making process for users when selecting products. Generally, users approach purchases with a preconceived price range or expectation in their minds. Regardless of their interest in an item, if its price surpasses their budgetary expectations, they are likely to forego the purchase. Owing to its critical influence on user decisions, the integration of item price into SBR holds the potential to significantly boost model performance and its applicability in real-world scenarios. By factoring in price considerations, SBR models can offer recommendations that are not only aligned with user interest but also their financial expectations, thereby enhancing user satisfaction. 


\subsection{Text Information}
{Text} information consists of item \emph{title} and \emph{description}. The item title serves as a succinct descriptive snippet, summarizing the item with a focus on its key features or selling points, such as size, capacity, and material. As to the description, it offers a more detailed account, giving potential buyers comprehensive insights into the items like usage instructions and support information. These textual elements depict item features from multiple angles, rendering them an invaluable asset for recommender systems. Due to their rich semantic content, text information has the potential to mitigate the data sparsity challenge of SBR. By utilizing the text with rich semantics rather than dull item IDs, neural models can generate more informative item representations, enhancing their effectiveness in the task. 

\subsection{Image Information}
{Image} information refers to the visual representations of items in the form of digital photographs or graphics. These images provide users with a clear and intuitive understanding of what the items look like, effectively showcasing key features such as color and style. This visual representation plays a crucial role in reducing the uncertainty that often accompanies online shopping. With the rich semantics, image information offers unique clues towards user preferences. As depicted in the second row of Figure~\ref{intro}, item images can be leveraged to discern a user's fine-grained preferences, \ie a Marvel fan, thereby significantly enhancing the effectiveness of recommendations. By harnessing the detailed insights gained from images, recommender systems are able to offer more precisely tailored suggestions, matching closely to individual user interests.

\subsection{Rating Information}
{Rating} serves as a quantifiable indicator of a user's overall attitude toward an item, typically expressed on a scale from 0 to 5. Generally, a higher rating indicates greater user preference for an item. Distinct from other types of information provided by manufacturers, ratings are uniquely generated by users themselves. This user-generated nature is particularly valuable for creating personalized recommendations. For instance, by analyzing users' explicit ratings on various items, recommender systems can identify and suggest items similar to items that have received high ratings, as opposed to items with lower ratings. The incorporation of ratings aligns recommendations more closely with the user's demonstrated preferences, resulting in a more tailored and satisfying user experience. 


\subsection{Review Information}
Review information consists of written comments provided by users who have purchased and used the items. Similar to ratings, reviews are user-generated content. However, they offer more detailed accounts of user preferences and item features, thanks to their narrative text format. These rich insights learned from reviews can significantly enhance the quality of recommendations. For instance, if a user expressed her dislike for tight clothing within a review, it would be a good alternative to offer her clothing with a loose style. Besides, we can also evaluate an item based on the reviews it received. If the majority of users rate an item as low-quality in their reviews, it suggests that the item might not be suitable for users seeking high-quality products. By leveraging such insights from reviews, recommender systems can make more informed and user-aligned recommendations.


\subsection{Behavior Information}
{Behavior} information refers to the specific types of user-item interactions, such as click, add-to-cart, or purchase. These distinct behavior types provide valuable insights into users' differing intentions and preferences regarding items. For instance, a user clicking on items may not necessarily indicate a strong preference or liking; it could simply be a part of casual browsing. On the other hand, a behavior like making a purchase is a strong indicator of user satisfaction with the item. The consideration of these diverse behaviors allows for a more fine-grained and deeper interpretation of user-specific intentions, significantly enhancing the effectiveness of SBR tasks.

\section{Research Progress}
In this section, we systematically summarize and discuss the research progress in the field of side information-driven SBR, with a particular emphasis on the types of information utilized in various studies. We highlight the techniques employed by different approaches. To provide a comprehensive understanding, in Table~\ref{approach}, we present an overview of the relevant models, information types each model incorporates, and the datasets on which these models have been applied.

\begin{table*}[th]
\small
\tabcolsep 0.08in 
  \centering

\begin{tabular}{c|c|c|c|    c|c|c|c|    c|c|c}
\toprule
\multirow{2}*{Models} & \multirow{2}*{Venues} & \multicolumn{8}{c|}{Types of Incorporated Information} & \multirow{2}*{Datasets} \\\cline{3-10}
& & {Category}    & {Brand}   & {Price}   & {Text}  & {Image}    & {Rating}  & {Review}  & {Behavior} & \\\hline

{P-RNN}    & {RecSys}    &{}    & {}   & {}   & {$\checkmark$}  & {$\checkmark$}    & {}  & {}  & {}    & {Private} \\  
 \hline
{FDSA}    & {IJCAI}    &{$\checkmark$}    & {$\checkmark$}   & {}   & {}  & {}   & {}    & {}  & {}    & {Amazon, Tmall} \\  
 \hline
{RNS}    & {IJCAI}    &{}    & {}   & {}   & {}  & {}   & {}    & {$\checkmark$}  & {}    & {Amazon} \\  
 \hline
 {MKM-SR}    & {SIGIR}    &{$\checkmark$}    & {$\checkmark$}   & {}   & {}  & {}    & {}  & {}  & {$\checkmark$}    & {JingDong} \\  
 \hline

 {S$^3$-Rec}    & {CIKM}    &{$\checkmark$}    & {$\checkmark$}   & {}   & {}  & {}    & {}  & {}  & {}    & {Amazon, LastFm, Yelp} \\  
 \hline
 {ESRM-KG}    & {WWW}    &{}    & {}   & {}   & {$\checkmark$}  & {}    & {}  & {}  & {}    & {Private} \\
  \hline
 {CoCoRec}    & {SIGIR}    &{$\checkmark$}    & {}   & {}   & {}  & {}    & {}  & {}  & {}    & {Beer, Taobao} \\  
 \hline
 {CBML}    & {CIKM}    &{$\checkmark$}    & {$\checkmark$}   & {}   & {}  & {}    & {}  & {}  & {}    & {Yoochoose, Diginetica} \\  
 \hline
 {NOVA}    & {AAAI}    &{$\checkmark$}    & {}   & {}   & {}  & {}    & {$\checkmark$}  & {}  & {}    & {MovieLens} \\  
 \hline

 {DIF-SR}    & {SIGIR}    &{$\checkmark$}    & {}   & {}   & {}  & {}    & {}  & {}  & {}    & {Amazon, Yelp} \\  
 \hline
 {MML} & {CIKM}    &{}    & {}   & {}   & {$\checkmark$}  & {$\checkmark$}    & {}  & {}  & {}    & {Private} \\  
 \hline
 {MGS}    & {SIGIR}    &{$\checkmark$}    & {$\checkmark$}   & {}   & {}  & {}    & {}  & {}  & {}    & {Diginetica, Tmall} \\  
 \hline
 {CoHHN}    & {SIGIR}    &{$\checkmark$}    & {}   & {$\checkmark$}   & {}  & {}    & {}  & {}  & {}    & {Amazon, Cosmetics, Diginetica} \\  
 \hline
 {UniSRec}    & {KDD}    &{$\checkmark$}    & {$\checkmark$}   & {}   & {$\checkmark$}  & {}    & {}  & {}  & {}    & {Amazon} \\  
 \hline
 {EMBSR}    & {ICDE}    &{}    & {}   & {}   & {}  & {}    & {}  & {}  & {$\checkmark$}    & {JingDong, Trivago} \\  
 \hline
 
 {NextIP}    & {CIKM}    &{}    & {}   & {}   & {}  & {}    & {}  & {}  & {$\checkmark$}    & {Tmall} \\   
 \hline
 {MBHT}    & {KDD}    &{}    & {}   & {}   & {}  & {}    & {}  & {}  & {$\checkmark$}    & {Retailrocket} \\   
 \hline
 {M2TRec}    & {RecSys}    &{$\checkmark$}    & {}   & {}   & {$\checkmark$}  & {}    & {}  & {}  & {}    & {Diginetica} \\  
 \hline
 
 {DETAIN}    & {WWW}    &{$\checkmark$}    & {$\checkmark$}   & {$\checkmark$}   & {}  & {}    & {$\checkmark$}  & {}  & {}    & {Amazon, LastFM} \\  
 \hline
  {DLFS-Rec}    & {RecSys}    &{$\checkmark$}    & {$\checkmark$}   & {}   & {}  & {}    & {}  & {}  & {}    & {Amazon} \\
 \hline
 {KMVG}    & {SIGIR}    &{$\checkmark$}    & {$\checkmark$}   & {}   & {}  & {}    & {}  & {}  & {}    & {Amazon, Cosmetics, Yelp} \\
 \hline
  {MMMLP}    & {WWW}    &{}    & {}   & {}   & {$\checkmark$}  & {$\checkmark$}    & {}  & {}  & {}    & {MovieLens} \\  
 \hline
  {M3SRec}    & {CIKM}    &{}    & {}   & {}   & {$\checkmark$}  & {$\checkmark$}    & {}  & {}  & {}    & {Amazon} \\
 \hline
 {MMSBR}    & {TKDE}    &{}    & {}   & {$\checkmark$}   & {$\checkmark$}  & {$\checkmark$}    & {}  & {}  & {}    & {Amazon} \\
 \hline
 {TASTE}    & {CIKM}    &{}    & {}   & {}   & {$\checkmark$}  & {}    & {}  & {}  & {}    & {Amazon, Yelp} \\
 \hline
 {MMSR}    & {CIKM}    &{}    & {}   & {}   & {$\checkmark$}  & {$\checkmark$}    & {}  & {}  & {}    & {Amazon} \\
 \hline
 {Recformer}    & {KDD}    &{$\checkmark$}    & {$\checkmark$}   & {}   & {$\checkmark$}  & {}    & {}  & {}  & {}    & {Amazon} \\
 \hline
 
 \hline

 \multicolumn{11}{c}{P-RNN~\cite{P-RNN}, FDSA~\cite{zhang@IJCAI2019}, RNS~\cite{RNS}, MKM-SR~\cite{meng@SIGIR2020}}\\
 \multicolumn{11}{c}{S$^3$-Rec~\cite{Zhou@CIKM2020}, ESRM-KG~\cite{liu@WWW2020}, CoCoRec~\cite{cai@SIGIR2021}, CBML~\cite{song@CIKM2021}}\\
 \multicolumn{11}{c}{NOVA~\cite{liu@AAAI2021}, DIF-SR~\cite{xie@SIGIR2022}, MML~\cite{Pan@CIKM2022}, MGS~\cite{lai@SIGIR2022}}\\
 \multicolumn{11}{c}{CoHHN~\cite{CoHHN}, UniSRec~\cite{hou@KDD2022}, EMBSR~\cite{yuan@ICDE2022}, NextIP~\cite{Luo@CIKM2022}}\\
 \multicolumn{11}{c}{MBHT~\cite{yang@KDD2022}, M2TRec~\cite{Shalaby@RecSys2022}, DETAIN~\cite{lin@WWW2022}, DLFS-Rec~\cite{Liu@RecSys2023}}\\
 \multicolumn{11}{c}{KMVG~\cite{Chen@SIGIR2023}, MMMLP~\cite{MMMLP}, M3SRec~\cite{Bian@CIKM2023}, MMSBR~\cite{MMSBR}}\\
 \multicolumn{11}{c}{TASTE~\cite{Liu@CIKM2023Text}, MMSR~\cite{Hu@CIKM2023}, Recformer~\cite{Li@KDD2023}}\\
\bottomrule
\end{tabular}
\vspace{-0.1in}
\caption{An overview of the relationships among current models, information types and datasets.}
\vspace{-0.1in}
\label{approach}
\end{table*}

\subsection{Time-Driven SBR}\label{timesbr}
{Time} is a key factor extensively investigated in SBR since its proposal~\cite{GRU4Rec,NARM}. Two primary ways to integrate this information into SBR are \textbf{temporal order} and \textbf{time interval}. 

For the temporal order manner, models only focus on the order of user-item interactions, emphasizing the relative position of interactions in a session~\cite{RepeatNet,LESSR}.  
These methods generally fall into two types: (1) applying RNN and its variants to capture sequential patterns within sessions~\cite{GRU4Rec,NARM,RepeatNet}; and (2) injecting an explicit position encoding into current algorithms including conventional KNN~\cite{Garg@SIGIR2019}, GNN~\cite{LESSR,FGNN,Li@CIKM2022} and self-attention~\cite{SASRec,STAMP}. 

In contrast, the time interval paradigm focuses on the specific duration between interactions, operating under the assumption that longer browsing times may indicate stronger user preferences~\cite{Li@WSDM2020,Wu@WWW2020}.
Relevant methods usually discretize continuous time intervals into one-hot encodings and rely on self-attention~\cite{Li@WSDM2020,Wu@WWW2020,Chen@CIKM2022} or contrastive learning~\cite{Tian@CIKM2022} to capture the time patterns. Another work~\cite{Dang@AAAI2023} employs time interval as an indicator to conduct uniform sequence augmentation. 


\subsection{Category-Driven SBR}\label{catesbr}
As the pioneering efforts incorporate item categories into this task, CoCoRec relies on self-attention layers to explore user in-category preferences for guiding model training~\cite{cai@SIGIR2021}. Some approaches model item ID sequences and category sequences simultaneously via GNNs~\cite{lai@SIGIR2022} or self-attention~\cite{Zhou@CIKM2020} for comprehensive session representation learning. 
Notably, existing methods typically convert item categories into one-hot encoding to serve as neural networks' inputs. However, these encoding approaches fail to capture the hierarchical associations among categories, resulting in a loss of valuable contextual information. 

\subsection{Brand-Driven SBR}\label{brandsbr}
None of the existing methods has specifically focused on item brand information in SBR. Instead, the item brand is typically considered in conjunction with other types of information in the task. Common practices involve combining item brand with category information, typically employing distinct self-attention layers to model item ID and side information separately for enriching session data~\cite{zhang@IJCAI2019,song@CIKM2021,Liu@RecSys2023}.  

\subsection{Price-Driven SBR}\label{pricesbr}
There are a few attempts to incorporate item prices into the SBR task~\cite{bipnet}. A representative attempt is the approach taken by CoHHN~\cite{CoHHN}, which builds a heterogeneous GNN encoding item ID, price, and categories to explicitly capture user price preferences. Given its unique role in the user decision process, we believe that much effort should be dedicated to introducing price factors into SBR.  

\subsection{Text-Driven SBR}\label{textsbr}
Various manners are presented to incorporate text into SBR, with the aim to leverage its rich semantics. For instance, ESRM-KG~\cite{liu@WWW2020} uses the keyword generation task to improve the recommendation task, whereas the former is designed to uncover shared intentions among items with distinct IDs. Some recent models~\cite{hou@KDD2022,Li@KDD2023} introduce pre-trained techniques in NLP domain into the user behavior modeling, transferring the language knowledge into the recommendation tasks. They employ text information to represent items and train models following the training-finetuning paradigm, achieving item and sequence representation learning. Due to its remarkable success in the NLP field, Transformer-based models are the most popular techniques used in these text-driven methods.


\subsection{Image-Involved Multi-modal SBR}\label{imagesbr}
Images are frequently combined with text in the recommendation tasks, which refers to multi-modal recommendations. In SBR, various methods are introduced to integrate image and text information for comprehensively mining user intents, such as simple weighted sum~\cite{P-RNN}, MLP~\cite{MMMLP}, self-attention mechanism~\cite{Pan@CIKM2022,Bian@CIKM2023,MMSBR} and GNN~\cite{Hu@CIKM2023}. 
However, the field of multi-modal SBR is still in its early stages, and several issues remain to be addressed, such as handling modality heterogeneity and noise, uncovering user preferences specific to each modality, and incorporating temporal information into the topic.

\subsection{Rating-Driven SBR}\label{ratingsbr}
While rating prediction was a primary research topic in the early stages of recommender systems, it has received little attention in the context of SBR. NOVA~\cite{liu@AAAI2021} presents a noninvasive fusion mechanism based on self-attention to merge ID, rating, and category information for user sequential behavior modeling. In their mechanism, fused information is used as Key and Query vectors, while the item ID is set as value vectors. Given the rating indicating the user's explicit attitude towards an item, we contend that this aspect deserves more investigation for improving SBR. 


\subsection{Review-Driven SBR}\label{reviewsbr}
Although possessing the huge potential to reveal fine-grained user preferences and item characteristics, user-item reviews receive little attention in SBR. RNS~\cite{RNS} creates item embeddings based on its reviews, employs an attention mechanism to learn a user's short- and long-term interests, and merges these interests to make final predictions. 


\subsection{Multi-behavior SBR}\label{behaviorsbr}
Current solutions for the emerging area of multi-behavior SBR typically build item sequences and behavior sequences separately. These sequences are then jointly analyzed to forecast the user's next actions, with an emphasis on the distinct purposes implied by various behaviors. Several advanced techniques are involved in these solutions, including RNN and GNN~\cite{meng@SIGIR2020,yuan@ICDE2022}, as well as hypergraph and Transformer~\cite{yang@KDD2022}. 


\subsection{Others}\label{othersbr}
In addition to the aforementioned information, some most recent efforts in SBR explore the use of more complex information. This includes not just item side information, but also the relations between different types of information. For instance, a ``\textit{belong-to}'' relation might indicate that an item belongs to a certain category or brand. Knowledge Graphs (KGs), which consist of entity nodes and relation edges, are adept at encoding these rich relations among various entities. 
Current methods~\cite{Wu@ICDE2023,Chen@SIGIR2023} incorporate KGs into SBR, thus, leveraging KG representation learning algorithms like TransE to derive item embeddings for making recommendations. 
However, one major challenge with KG-based SBR is the requirement for a well-structured KG, which often necessitates extensive domain knowledge and significant labor to construct and maintain. This issue seriously limits the development of KG in SBR.


\section{Open Challenges and Future Directions}
In this section, we identify and discuss several open research directions, setting the stage for future breakthroughs and advancements in the field.

\paragraph{(1) Strengthen the utilization of current information in SBR.}
As discussed earlier, each type of information, ranging from time and price to images and reviews, offers useful perspectives on user preferences. These diverse data types are instrumental in uncovering specific user interests and enabling the provision of personalized services.  Unfortunately, as highlighted in the section of Research Progress, certain types of information, such as brand, price, and reviews, have not been extensively explored or utilized in SBR. In light of this, more efforts should be dedicated to incorporating such less-exploited information into SBR, employing their unique utility to facilitate understanding of user intent.

\paragraph{(2) Joint incorporation of various information into SBR.} 
Current methods in SBR, as outlined in Table~\ref{approach}, typically focus on leveraging only a subset of available information. However, none have yet explored the potential of jointly employing these diverse information sources.
The integrated utilization of all varied information promises to not only enrich session data but also to provoke unique insights, leading to a more comprehensive understanding of user behaviors. For example, the integration of time information with color attributes derived from item images could uncover seasonal trends in user preferences, such as a predilection for lighter colors in summer apparel and darker tones for winter clothing. It is a promising direction to further alleviate the sparsity issue in SBR and improve its effectiveness.

\paragraph{(3) Side information enhanced cold-start SBR.}
The cold-start problem is a persistent challenge in recommender systems, where the systems fall short of offering suggestions about new items. However, little attention has been paid to solving this issue in SBR. Actually, the incorporation of side information presents a promising solution for this issue. By leveraging various information, such as images and text, models can capture rich semantics about new items, effectively alleviating cold-start issues.

\paragraph{(4) Side information driven explainable SBR.}
Providing explanations for recommendation results can significantly boost user satisfaction and trust in the system. Side information serves as a rich source for uncovering user specific interests, such as identifying a user’s preference for Marvel as presented in \fig~\ref{intro}, contributing to generating insightful explanations. As a result, leveraging side information to create explainable SBR would be an important direction for future research.

\paragraph{(5) Side information enhanced cross-field SBR.}
Cross-field recommendation seeks to transfer knowledge from a source domain, rich in user-item interactions, to a target domain where such interactions are sparse, enhancing the user experiences in the target field. The item side information is ideally suited for the transferring operations since user fine-grained preferences about such information are relatively stable and independent from user-item interaction data. However, few efforts have been dedicated to the direction.

\paragraph{(6) LLM-based SBR with side information.}
Exhibiting promising capabilities in various tasks, Large Language Models (LLMs) have been increasingly explored in the recommendation field. However, their initial attempts into this domain have not been as successful as anticipated. We contend that while LLMs excel at handling tasks involving modalities rich in semantics, like text or images, they struggle to adapt to current ID-based recommendations. This difficulty is particularly significant in SBR, where sessions often consist of limited user actions. Side information discussed in this survey, especially item text and images, presents a significant opportunity to harness the full potential of LLMs, which warrants extensive and in-depth investigation.

\paragraph{(7) Side information based benchmarks for SBR.}
As emphasized in previous sections, datasets are a vital resource in our researched topic. However, as presented in Table~\ref{datasets}, currently available datasets do not encompass all types of information, which presents a significant challenge for comparative analysis in this field. For instance, one method may focus on ``Behavior'' using Cosmetics, while another one might concentrate on ``Images'' via Amazon. Without a dataset including both ``Behavior'' and ``Images'', it becomes impossible to compare the performance of these two methods effectively. Therefore, we advocate for the creation and availability of more comprehensive and high-quality datasets to promote the development of this promising direction.

\section{Conclusions}
The session-based recommendation is a hot-spot research topic due to its substantial application value in offering personalized services for anonymous users based on their limited behaviors. 
In recent times, there has been a surge in efforts to integrate side information into SBR as a means to address its inherent issues of data sparsity.
This survey elaborates on side information-driven SBR from a data-centric perspective, including benchmarks, data characteristics and utility, progress, and future directions. We sincerely hope that our insights and analyses will serve as a valuable resource for researchers and practitioners, fostering further innovations and developments in this vibrant topic.


\newpage

\bibliographystyle{named}
\bibliography{main}

\end{document}